\documentclass[fleqn,usenatbib]{mnras}

\usepackage[T1]{fontenc}

\DeclareRobustCommand{\VAN}[3]{#2}
\let\VANthebibliography\thebibliography
\def\thebibliography{\DeclareRobustCommand{\VAN}[3]{##3}\VANthebibliography}

\usepackage{graphicx}	% Including figure files
\usepackage{amsmath}	% Advanced maths commands
\usepackage{amssymb}	% Extra maths symbols
\usepackage{lscape}
\usepackage{ulem}
\usepackage{tabularx}
\usepackage[dvipsnames]{xcolor}
\usepackage{soul}

\usepackage{caption}
\usepackage{subcaption}

\graphicspath{{images/}}
\title[Signature of a chemical spread in M37]{Signature of a chemical spread in the open cluster M37}

\author[M. Griggio et al.]{M. Griggio$^{1,2}$\thanks{E-mail: massimo.griggio@inaf.it},
M. Salaris$^{3,4}$,
S. Cassisi$^{4,5}$, 
A. Pietrinferni$^{4}$ and
L. R. Bedin$^{2}$
\\
$^{1}$Dipartimento di Fisica, Universit\`a di Ferrara, Via Giuseppe Saragat 1, Ferrara I-44122, Italy\\
$^{2}$INAF - Osservatorio Astronomico di Padova, Vicolo dell'Osservatorio 5, Padova I-35122, Italy\\
$^{3}$Astrophysics Research Institute, Liverpool John Moores University, 146 Brownlow Hill, Liverpool L3 5RF, UK\\
$^{4}$INAF - Osservatorio Astronomico di Abruzzo, Via M. Maggini, I-64100 Teramo, Italy\\
$^{5}$INFN - Sezione di Pisa, Largo Pontecorvo 3, 56127 Pisa, Italy
}

\date{Accepted 2022 August 31. Received 2022 August 31; in original form 2022 July 11}

\pubyear{2021}

\usepackage{newtxtext,newtxmath}

\begin{document}
\label{firstpage}
\pagerange{\pageref{firstpage}--\pageref{lastpage}}
\maketitle

\begin{abstract}
Recent {\it Gaia} photometry of the open cluster M37 have disclosed the existence of an 
extended main sequence turn off --like in Magellanic clusters younger than about 2~Gyr--  
and a main sequence that is broadened in colour  
beyond what is expected from the photometric errors, at magnitudes well below the region of the 
extended turn off, where neither age differences nor rotation rates 
(the candidates to explain the extended turn off phenomenon) are expected to play a role. 
Moreover, not even the contribution of unresolved binaries can fully explain the observed broadening.
We investigated the reasons behind this broadening by making use of synthetic stellar populations and differential colour-colour diagrams using a combination of {\it Gaia} and {\it Sloan} filters.
From our analysis we have 
concluded that the observed colour spread in the {\it Gaia} colour-magnitude diagram can be reproduced 
by a combination of either a metallicity spread $\Delta \rm[Fe/H] \sim 0.15$ plus a differential reddening across the face of the cluster spanning a total 
range $\Delta E(B-V) \sim 0.06$, or a spread of the initial helium mass fraction 
$\Delta Y\sim0.10$  plus a smaller range of reddening $\Delta E(B-V) \sim 0.03$.
High-resolution differential abundance determinations of a sizeable sample of cluster stars 
are necessary to confirm or exclude the presence of a metal abundance spread.
Our results raise the possibility that also individual open clusters, like globular clusters and massive star clusters,
host stars born with different initial chemical compositions.

\end{abstract}


%%%%%%%%%%%%%%%%%%%%%%%%%%%%%%%%%%%%%%%%%%%%%%%%%%

%%%%%%%%%%%%%%%%% BODY OF PAPER %%%%%%%%%%%%%%%%%%

\section{Introduction}

The study of star clusters has been and still is one of the main sources of 
information about stars and galaxies. 
Photometric as well as spectroscopic observations allow us to determine a cluster's kinematics, its 
distance, age, chemical composition, dynamical status and the detailed colour and magnitude 
distribution of its stars, all pieces of information that set strong constraints
on astrophysical models of galaxy and stellar evolution. 

In these studies a first crucial step 
is the determination of the membership probability of the observed
stars, to disentangle actual cluster members from neighbouring field stars
not bound to the cluster.
Recently, \citet{gb2022} developed a new formalism to compute the astrometric
membership probabilities for sources in star clusters,
and applied their technique to the Galactic open cluster NGC~2099 (M37) 
using {\it Gaia} Early Data Release 3 (EDR3) data.

This cluster has an age of about 500~Gyr, a metallicity 
around Solar, and 
has been the subject over the years of several investigations regarding 
its distance, age and dynamical status \citep[e.g.][]{mermilliod, sagar, joshi}, 
searches for variable 
\citep[e.g.,][]{kiss, kang} and peculiar stars \citep{paunzen03}, studies 
of rotation and photometric activity of its low-mass stellar population
\citep[e.g.][]{mess, chang}, investigations of its white dwarf initial-final mass relation 
\citep[e.g.][]{kalirai05, cumm15, cummings2016}, and its white dwarf cooling sequence 
\citet{kalirai01}.
More recently \citet{cordoni18} employed photometry and proper
motions from {\it Gaia} Data Release~2 to reveal the presence of an extended main sequence (MS) 
turn off (TO) in the colour-magnitude diagram (CMD) of M37 
(and a few other Galactic open clusters), qualitatively similar to what found in Magellanic Cloud clusters younger than about 2~Gyr.

In this paper we have exploited the accurate {\it Gaia} EDR3 CMD provided by \citet{gb2022} 
which shows a MS broadened not only around the TO 
\citep[as found by][]{cordoni18}, but also in the lower mass regime (i.e. when $G\gtrsim15$).
We will show that this broadening of the lower MS is not due just to photometric errors, and also that differential reddening plays only a minor role.
A moderate spread of metallicity or helium appear to be the main culprit, adding an unexpected new twist to our evolving views about star clusters and their formation.
To state it more clearly, this work will not deal with the MSTO phenomenon, but will focus on the part of the MS which is not affected to both rotation and age effects.

The paper is organised as follows.
In Section~\ref{CMD} we present the {\it Gaia} photometry; the auxiliary {\it Sloan} photometry which we used to study the lower MS broadening is briefly described in Section~\ref{ugiSch}.
This is followed by Section~\ref{broad} which presents our detailed analysis 
of the broadening of the lower MS. A section with a summary and conclusions brings the paper to a close.

\newpage

%%%%%%%%%%%%%%%%%%%%%%%%%%%%%%%%%%%%%%%%%%%%%%%%
%%%%%%%%%%%%%%%%%%%%%%%%%%%%%%%%%%%%%%%%%%%%%%%%
%%%%%%%%%%%%%%%%%%%%%%%%%%%%%%%%%%%%%%%%%%%%%%%%
%%%%%%%%%%%%%%%%%%%%%%%%%%%%%%%%%%%%%%%%%%%%%%%%
\section{The \textit{Gaia} colour-magnitude diagram}
%%%%
\label{CMD}
\begin{figure}
    \centering
    \includegraphics[width=\columnwidth]{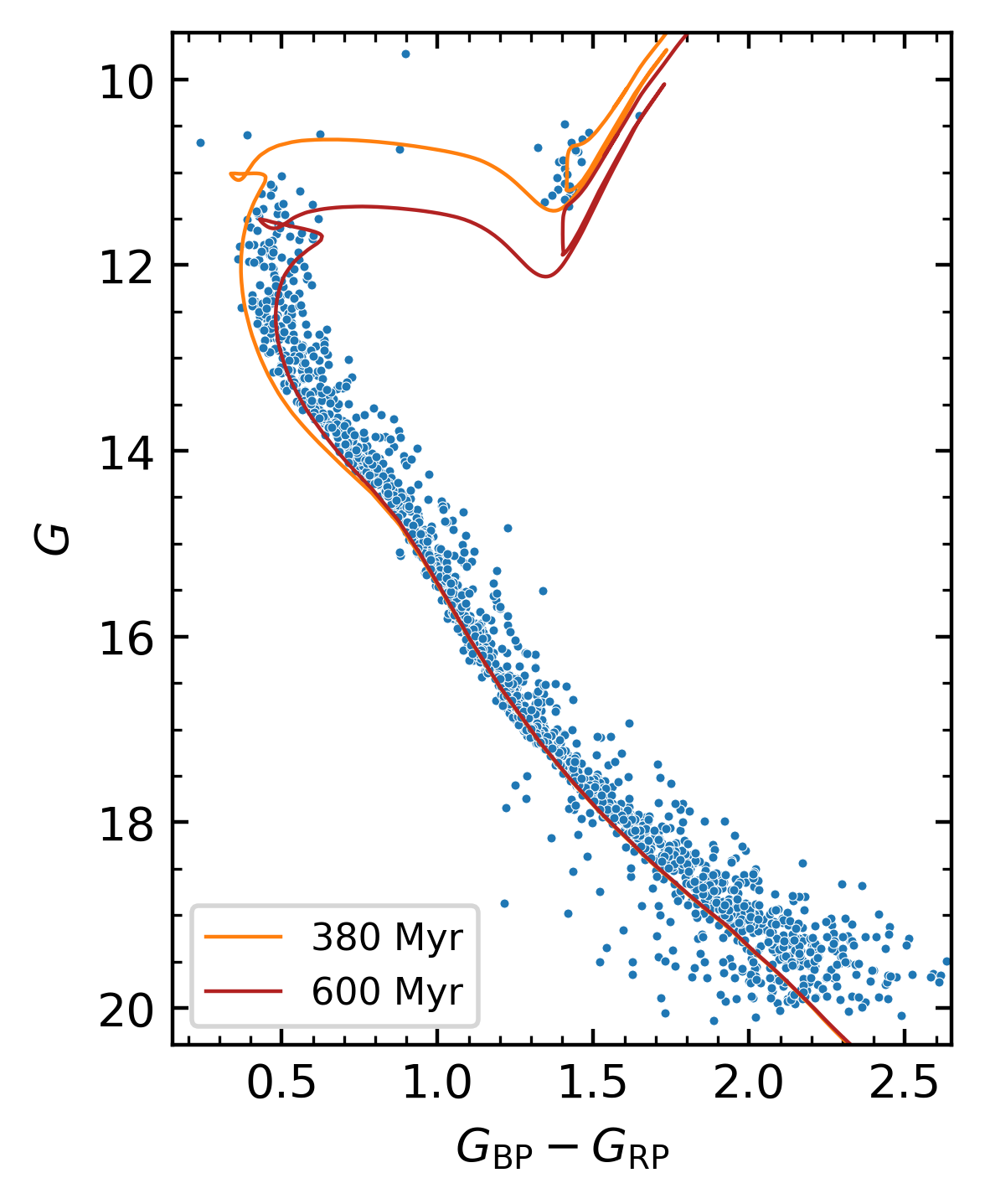}
    \caption{{\it Gaia} EDR3 CMD of M37 with superimposed two non-rotating 
    BaSTI-IAC isochrones with $\rm [Fe/H]=0.06$ and the two labelled ages, 
    bracketing the extended TO region 
    (see text for details).}
    \label{fig:cmd_iso}
\end{figure}

Figure~\ref{fig:cmd_iso} displays the CMD of M37 from {\it Gaia} EDR3, as 
obtained by \citet{gb2022}. The diagram clearly exhibits an extended TO region, a red clump of core He-burning stars (around $G \sim 11$ and $(G_{\mathrm{BP}}-G_{\mathrm{RP}}\sim1.4$), and a  
MS with a parallel sequence of unresolved binaries with mass ratio greater than $\sim 0.6$-0.7.

By employing a cluster distance equal to 1.5\,$\pm$\,0.1\,kpc \citep[determined from {\it Gaia} EDR3 parallaxes by][]{gb2022}, we qualitatively checked the general consistency of theoretical 
isochrones with the  
cluster CMD. More specifically, we adopted as reference the non-rotating Solar-scaled BaSTI-IAC isochrones \citep{hidalgo18} that include convective core overshooting. 

Figure \ref{fig:cmd_iso} displays two $\rm [Fe/H]=0.06$ (corresponding to an initial metallicity $Z=0.0172$
and helium mass fraction $Y=0.269$) isochrones with ages equal to 380 
and 600~Myr respectively, 
matched to the blue edge\footnote{The blue edge of the observed CMD is determined as described in Sect.~\ref{broad}. It will be clearer 
in Sect.~\ref{broad} why we match the blue edge of the unevolved MS in this qualitative comparison.} of the observed 
unevolved MS between $G \sim 15$ and $G\sim 17.5$ for a distance equal to 1450\,pc. 
We took into account the extinction by employing the 
extinction law for the {\it Gaia} filters given by the {\it Gaia} collaboration\footnote{\url{https://www.cosmos.esa.int/web/gaia/edr3-extinction-law}} and derived a reddening $E(B-V)=0.28$. The two ages employed in this comparison approximately bracket the brighter 
and lower limit of the cluster's extended TO region.

The reddening is consistent with the broad range of values found in the literature 
\citep[$E(B-V)$ between $\sim 0.23$ and $\sim 0.35$, see, e.g.,][]{piatti, hartman, joshi}, and the metallicity chosen for the isochrones is 
also consistent with the values of [Fe/H] measured with high- and low-resolution spectroscopy 
\citep[see, e.g.][]{pancino, marshall, netopil}, typically in the range between about Solar and $\sim +0.10$~dex.

In this comparison we employed non-rotating isochrones, and the extended TO region is bracketed by assuming a total 
age spread of about 220~Myr \citep[see also][]{cordoni18}. 
Another possibility \citep[likely the one to be preferred, see e.g.][]{cordoni18, bastian} is that the extended TO 
is caused by the presence of stars with approximately the same age but with a range of  
initial rotation rates.

Irrespective of the reasons for the appearance of the extended TO, the cluster MS with $G$ larger 
than $\sim 15$-15.5 is predicted to be insensitive to either an age spread (stars 
in this magnitude range are still essentially on their 
zero age MS location at these ages) and a spread of initial rotation rates. This is because, at the cluster's metallicity, stars in this magnitude range have masses below $\sim 1.25$-1.2\,$M_{\odot}$, with convective envelopes thick enough
for magnetic braking to efficiently spin them down enough and suppress the effects of rotation (on the hydrostatic equilibrium and chemical mixing) that cause the MS broadening 
\citep[see, e.g,][]{rotgen, rotmesa}. 
We could also test empirically that rotation does not play a role in 
the colour spread of the lower MS, by cross-correlating the measurements of rotational periods of M37 MS stars by 
\citet{chang} with our {\it Gaia} photometry. We ended up with a sample of more than 150 member stars with $G$ between $15.5$ and $17$ and periods centred around $\sim 6$ days, that do not show any correlation with the  $G_{\mathrm{BP}}-G_{\mathrm{RP}}$ colour at a given $G$ magnitude.

We have therefore studied the thickness of the MS for 
$G$ magnitudes larger than 15.5, to avoid the impact of the extended TO phenomenon. As 
faint limit we considered 
$G=17$ (corresponding to a stellar mass $\sim0.95$\,$M_{\odot}$), because at larger magnitudes the membership probability is more uncertain \citep[see][]{gb2022},
leading to a contamination of the CMD by non-member stars.

In the standard assumption that open clusters host single-metallicity populations, 
the observed colour width of the MS in the selected magnitude range 
is expected to be set by 
the photometric error, the presence of unresolved binaries with a range of values of the mass ratio $q$, and a possible differential reddening across 
the face of the cluster.
To verify this expectation, we have produced a synthetic CMD of the MS in 
this $G$-magnitude range (we will use the term lower-MS from now on, to denote 
this specific magnitude range along the cluster MS) 
for the case of single stars all with the same initial metallicity, as 
described below.

\begin{figure}
    \centering
    \includegraphics[width=\columnwidth]{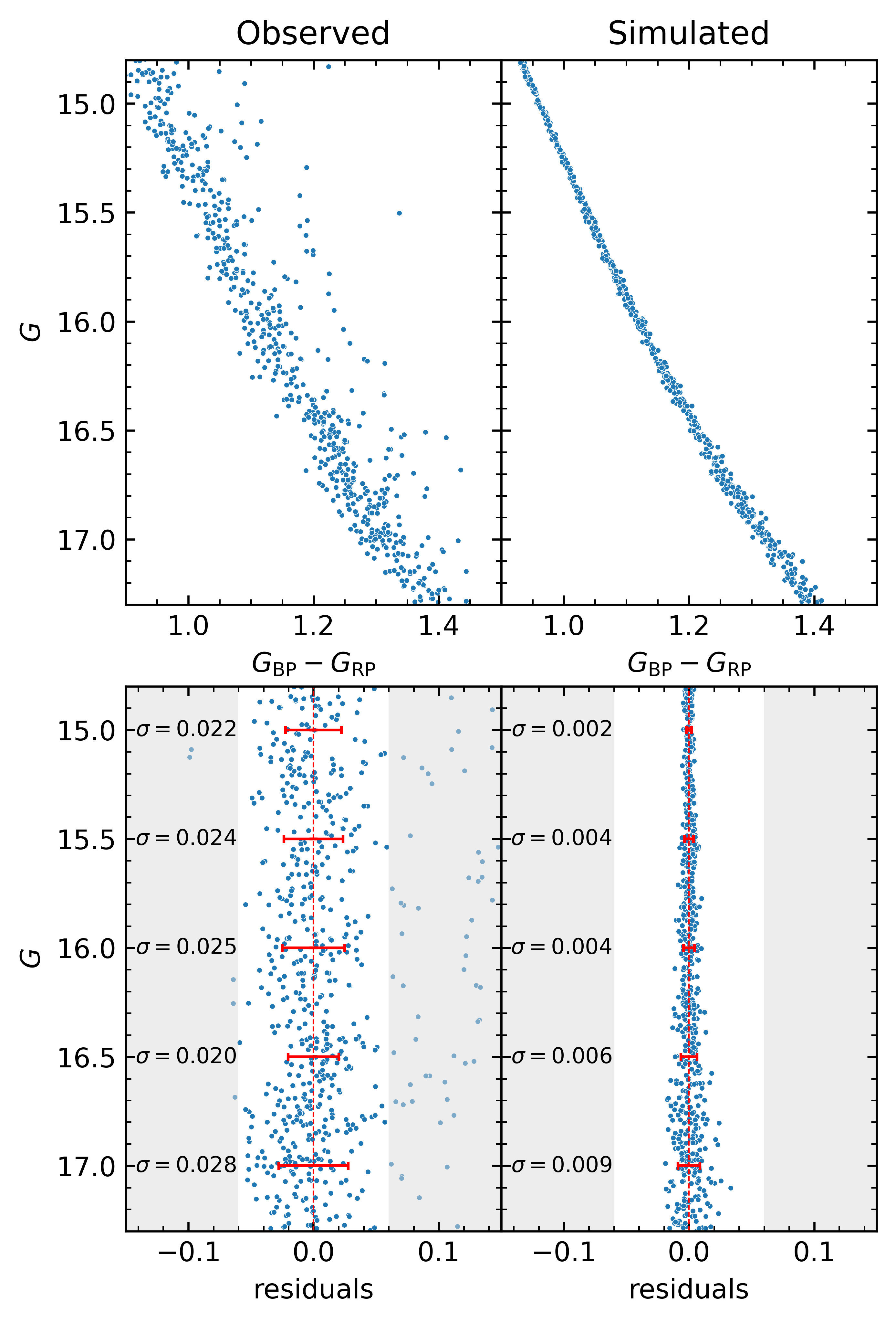}
    \caption{{\sl Top panels:} Observed (left) and synthetic (right) CMDs of the cluster lower MS. {\sl Bottom panels:} Colour residuals around the lower MS fiducial lines 
    of the observed (left) and synthetic (right) CMDs, as a function of the $G$ magnitude. The white area encloses the stars employed in the calculation 
    of the 1$\sigma$ values of the dispersion of the colour residuals reported in 
    the two panels (see text for details).}
    \label{fig:cmd_err}
\end{figure}

We have defined an observed fiducial line by partitioning the CMD into 0.5~mag wide $G$-magnitude bins, and interpolated with a quadratic spline the median points of the magnitude and colour number distributions within each bin. We have then uniformly distributed synthetic stars along the fiducial, by adding photometric errors randomly sampled from a Gaussian distribution with zero mean and a standard deviation equal to the median error at the corresponding 
$G$-magnitude (individual errors are taken from the {\it Gaia} EDR3 photometry). 
The top panels of Fig.~\ref{fig:cmd_err} show the observed cluster CMD (left) and the synthetic 
CMD described above (right) for the relevant MS region, 
while the bottom panels display the colour residuals around the fiducial line 
as a function of $G$. We also report the values of the dispersion of the colours around the fiducial values at different magnitudes in both CMDs, calculated as the $68.27^{\rm th}$-percentile of the distribution of the residuals around zero. 

In the calculation of the dispersion of the residuals for the observations we have neglected objects whose place in the CMD is compatible with the 
position of unresolved binaries with mass ratio $q > 0.6$ (as determined using our isochrones). 
Even after excluding these objects, the lower panels of Fig.~\ref{fig:cmd_err} show clearly that the synthetic stars are much more narrowly distributed around the fiducial line, when   
compared to the observations. 

Within the standard assumptions described before, the broader colour range 
spanned by the observed CMD at a given value of $G$ might be ascribed to the presence of unresolved 
binaries with $q$ lower than 0.6, plus possibly the effect of differential reddening. 
To understand whether this is the case, we took advantage of an auxiliary photometry in the {\it Sloan} $ugi$ filters, 
described in the following section, that we 
combined with the {\it Gaia} data as discussed in the Sect.\,\ref{broad}.

%%%%%%%%%%%%%%%%%%%%%%%%%%%%%%%%%%%%%%%%%%%%%%%%
%%%%%%%%%%%%%%%%%%%%%%%%%%%%%%%%%%%%%%%%%%%%%%%%
%%%%%%%%%%%%%%%%%%%%%%%%%%%%%%%%%%%%%%%%%%%%%%%%
%%%%%%%%%%%%%%%%%%%%%%%%%%%%%%%%%%%%%%%%%%%%%%%%

\section{The \textit{Sloan} colour-magnitude diagram}
\label{ugiSch}

Our adopted {\it Sloan} photometry is taken from the catalogue presented and described in \cite{griggio22}.
Briefly, the data has been collected with the Schmidt 67/92\,cm telescope 
in Asiago (Italy), and the photometry extracted with
a version of the \texttt{KS2} software by \cite{2008AJ....135.2055A} suitably modified to deal with the Schmidt data
and wide field mosaics. We selected only the sources with the quality flag \texttt{pho\_sel} equal to one, to reject sources with poor photometry.

\begin{figure}
    \centering
    \includegraphics[width=\columnwidth]{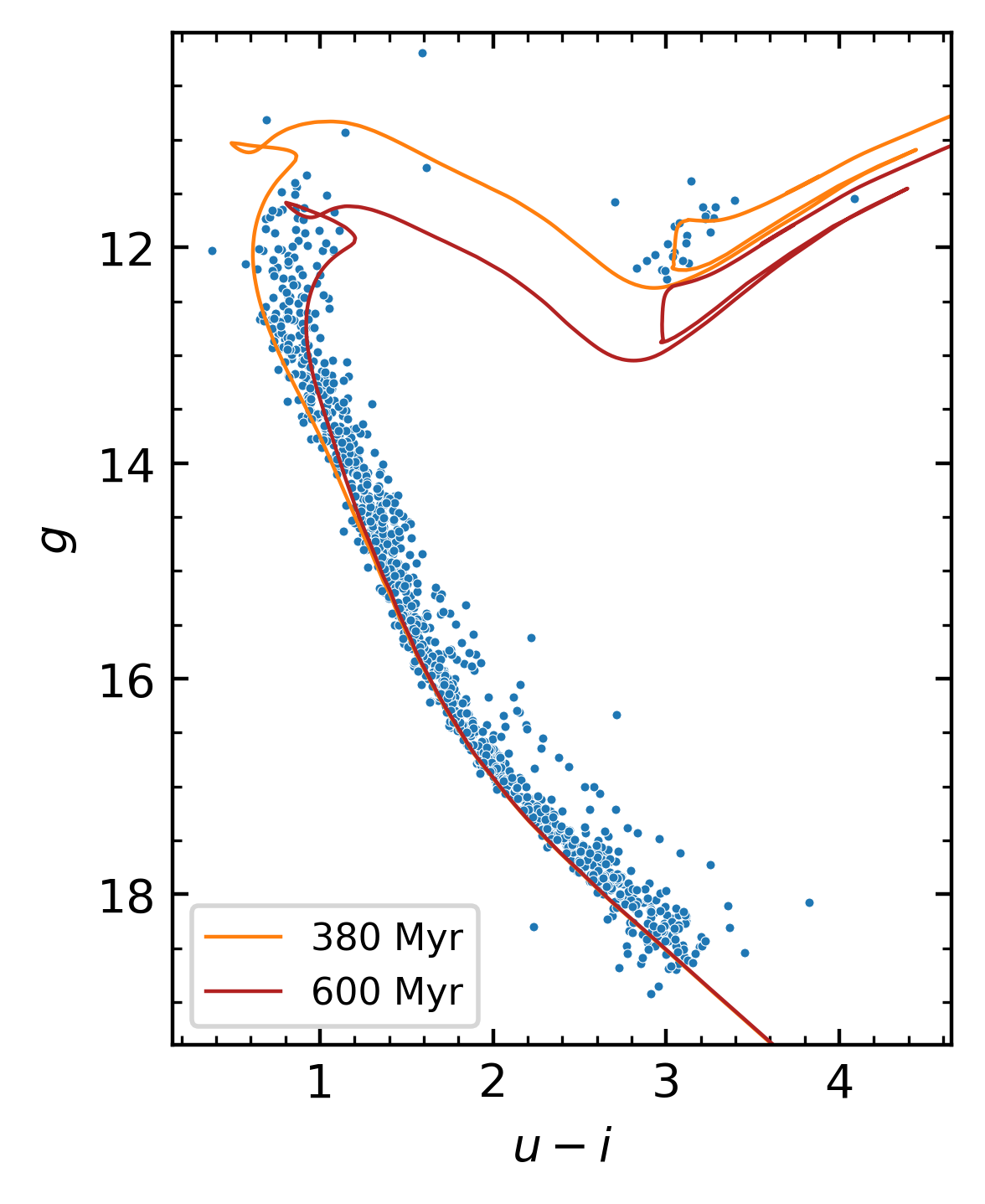}
    \caption{As Fig.~\ref{fig:cmd_iso} but in the $g$-$(u-i)$ CMD.}
    \label{fig:cmd_iso_sl}
\end{figure}

Figure~\ref{fig:cmd_iso_sl} shows the $g$-$(u-i)$ CMD for cluster's members identified with {\it Gaia}, 
together with the same isochrones of Fig.~\ref{fig:cmd_iso}, which are compared to the observations by employing the same distance and $E(B-V)$ values of the match to the $Gaia$ CMD.

The effect of interstellar extinction on the magnitudes of the theoretical isochrones has been included using the extinction ratios $A_{\lambda}$/$A_V$ from the NASA/IPAC infrared science archive\footnote{\url{https://irsa.ipac.caltech.edu/applications/DUST/}} for the filters 
$u$, $g$ and $i$\footnote{These ratios are $A_u/A_V=4.239$, $A_g/A_V=3.303$, and $A_i/A_V=1.698$}.
Notice that the comparison of the observed CMD with the isochrones is completely consistent with the results for the corresponding CMD in the {\it Gaia} filters.

%%%%%%%%%%%%%%%%%%%%%%%%%%%%%%%%%%%%%%%%%%%%%%%%
%%%%%%%%%%%%%%%%%%%%%%%%%%%%%%%%%%%%%%%%%%%%%%%%
%%%%%%%%%%%%%%%%%%%%%%%%%%%%%%%%%%%%%%%%%%%%%%%%
%%%%%%%%%%%%%%%%%%%%%%%%%%%%%%%%%%%%%%%%%%%%%%%%

\section{The broadening of the lower MS}
\label{broad}

To investigate in detail the broadening of the lower MS in the range $15.5 \leq G \leq 17$ (we have a total of 387 stars in this magnitude range) we combined the photometry in the {\it Gaia} filters with the corresponding $u$ and $i$ magnitudes to build a differential colour-colour diagram, as follows.

As a first step we have defined a MS blue fiducial line in both the $G$-$(G_{\mathrm{BP}}-G_{\mathrm{RP}})$ and $G$-$(u-i)$ diagrams, by partitioning  
the data into $G$-magnitude bins 0.2~mag wide. For each bin we have 
first performed a $3\sigma$-clipping around the median values of the magnitudes and colours, and then we have calculated 
a representative colour corresponding to the $10^{\rm th}$-percentile of the colour distribution, and the mean $G$-magnitude.
We have finally interpolated with a linear spline among these 
pairs of colours and magnitudes determined for each bin, to calculate the blue fiducial line for each diagram.

\begin{figure}
    \centering
    \includegraphics[width=\columnwidth]{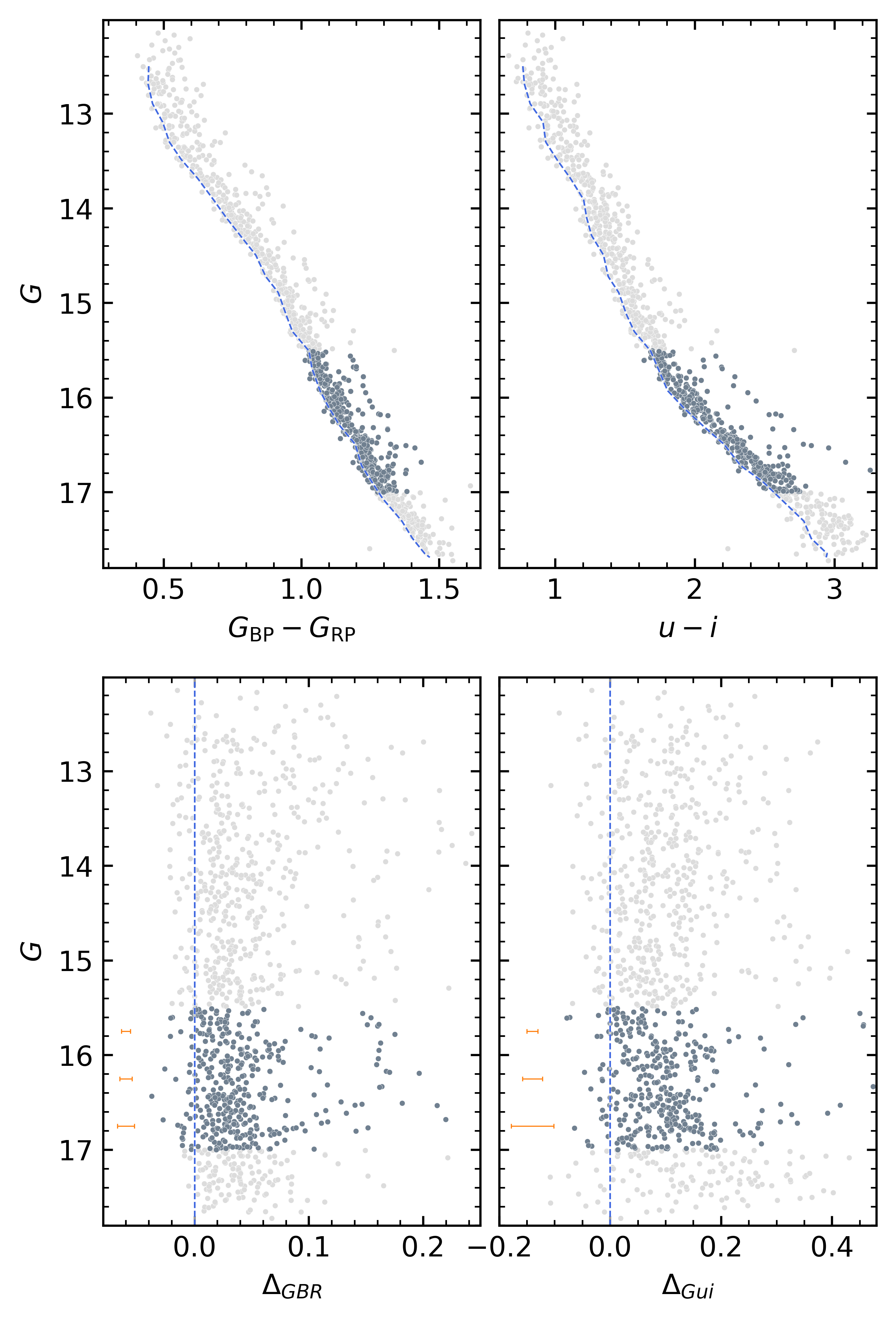}
    \caption{{\sl Top panels:} The cluster $G$-$(G_{\mathrm{BP}}-G_{\mathrm{RP}})$ and $G$-$(u-i)$ diagrams. The MS blue fiducials are displayed as dashed lines. Stars in the $G$ magnitude range of interest are displayed in a darker grey shade.
    {\sl Bottom panels:} The $G$-$\Delta_{GBR}$ and $G$-$\Delta_{Gui}$ diagrams 
    (see text for details). On the left of each panel we display the median $\pm 1 \sigma$ 
    colour error at three representative $G$ magnitudes.}
    \label{fig:ddres}
\end{figure}

For each observed star we have then computed, in the $G$-$(G_{\mathrm{BP}}-G_{\mathrm{RP}})$ and $G$-$(u-i)$ diagrams,
the difference between its colour and the corresponding value of the blue fiducial at the star $G$ magnitude. We notice here that the error on the $G$ magnitudes of the individual stars is on the order of 0.001~mag.
We denote these quantities as $\Delta_{GBR}$ and $\Delta_{Gui}$ respectively (see Figure \ref{fig:ddres}).

\begin{figure}
    \centering
    \includegraphics[width=\columnwidth]{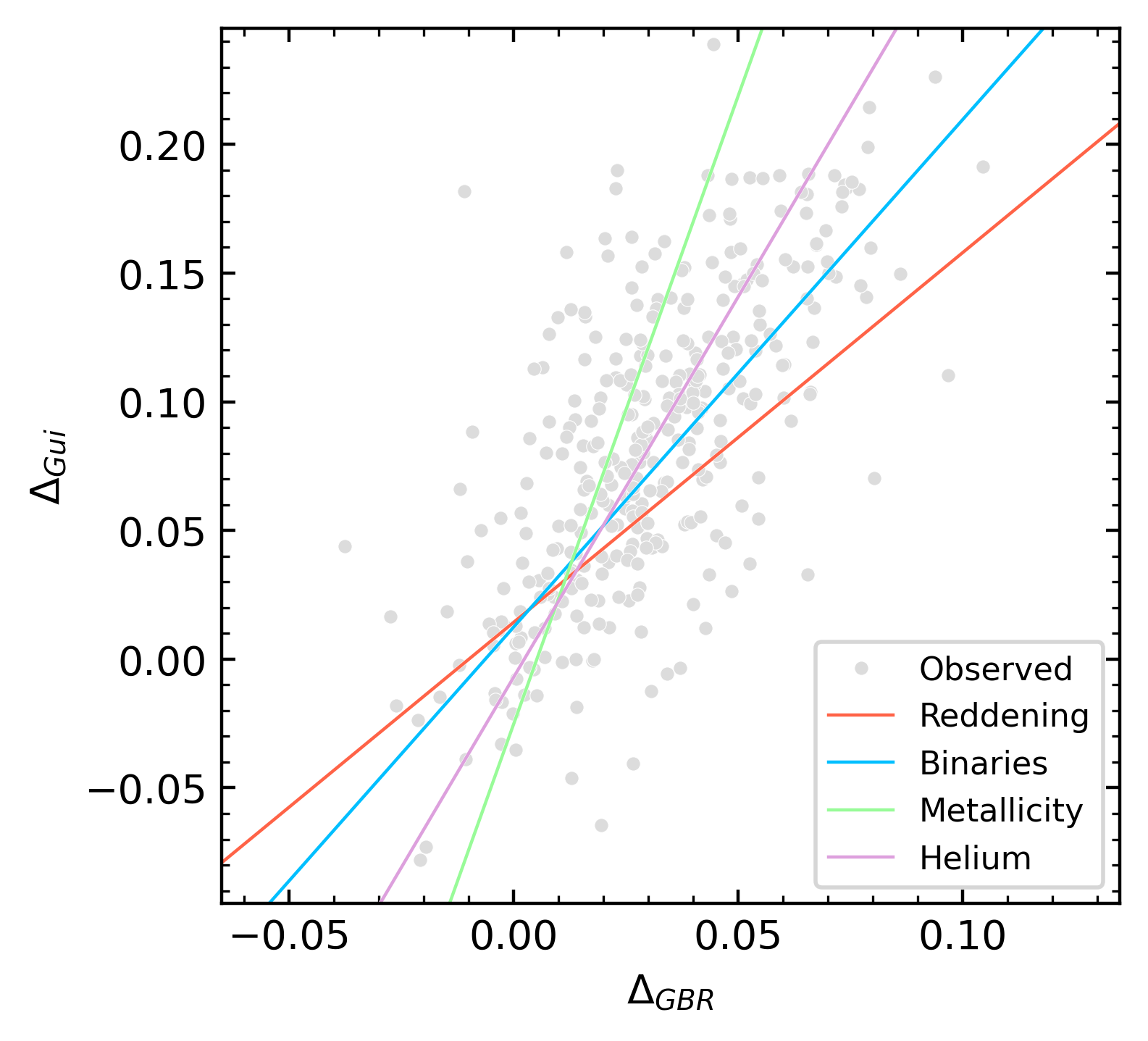}
    \caption{$\Delta_{GBR}$-$\Delta_{Gui}$ diagram for the lower 
    MS stars, after excluding sources whose colours are consistent with unresolved binaries with mass ratio $q\gtrsim 0.6$. The four straight lines display the the direction along which stars 
    would move due to the effects of differential reddening, unresolved binaries with  
    $q\lesssim 0.6$, spread of initial metallicity, and spread of initial helium abundance (see text for details).}
    \label{fig:obs_red}
\end{figure}

We finally plotted these colour differences in a 
$\Delta_{GBR}$-$\Delta_{Gui}$ diagram shown in 
Fig.~\ref{fig:obs_red}, after excluding the relatively small number of 
sources whose colours are 
consistent with unresolved binaries with $q \gtrsim 0.6$,
which are clearly separated from the 
bulk of the MS in the $Gaia$ CMD.
The lower MS stars are distributed along a clearly defined sequence 
with origin around the coordinates (0,0) --that correspond to stars lying on the blue fiducials-- and extended towards 
increasingly positive values (corresponding to stars progressively redder than the fiducials) with $\Delta_{Gui}$ increasing faster than $\Delta_{GBR}$.
If the colour spreads are due to random photometric 
errors only, stars would be distributed without a correlation 
between $\Delta_{Gui}$ and $\Delta_{GBR}$, as we have 
verified by calculating a synthetic sample of cluster 
stars including only the photometric errors, 
as described in more detail below.

This diagram allows us to exclude that differential reddening is the 
main reason for the broadening of the cluster lower MS. 
Figure~\ref{fig:obs_red} shows together with the data also 
the direction of the reddening vector, and  
we can see that differential reddening would move stars at a different angle (shallower) 
compared to the observed trend. We have also tried 
the alternative $A_{\lambda}/A_V$ extinction ratios for the $u$ and $i$ filters
presented in \citet{2013MNRAS.430.2188Y} and
\citet{tian}, but the slope of the reddening vector in this diagram hardly changes.

To study in more detail the origin of the 
distribution of points in this $\Delta_{GBR}$-$\Delta_{Gui}$ diagram, we 
have used the theoretical isochrones of Figs.\,\ref{fig:cmd_iso} and \ref{fig:cmd_iso_sl} to calculate synthetic samples of lower MS 
stars as follows. We have considered 600~Myr, $\rm [Fe/H]=0.06$ isochrones 
(the choice of age is irrelevant in this magnitude range) as a reference, 
and drawn randomly 50000 values of the stellar mass in the range covering the MS, according to a power law mass function with exponent equal to $-$2.3\footnote{This choice of the mass function provides a good match to the distribution of stars as a function of the $G$ magnitude in 
the magnitude range of interest.}. By interpolating along the 
isochrones we determined the ${\it Gaia}$ and 
$Sloan$ magnitudes of these synthetic objects.
We have then considered the contribution of unresolved binaries with $q < 0.6$ 
by extracting randomly (with a uniform probability distribution) for each  
synthetic star the value of the mass ratio $q$ to the secondary, to 
calculate the mass of the unresolved companion. The magnitudes of the companion 
in the {\it Gaia} and {\it Sloan} $u$ and $i$ filters 
are then derived as described before, and the fluxes of the two components  
added to determine the total magnitudes of the corresponding unresolved system. 
To these magnitudes we added the distance modulus and extinction derived from the fit in Fig.~\ref{fig:cmd_iso} 
and applied random 
Gaussian photometric errors by considering the median 1$\sigma$ errors of the observations 
at the $G$ magnitude of the synthetic star.

\begin{figure*}
     \centering
     \begin{subfigure}{0.495\textwidth}
        \centering
        \includegraphics[width=\textwidth]{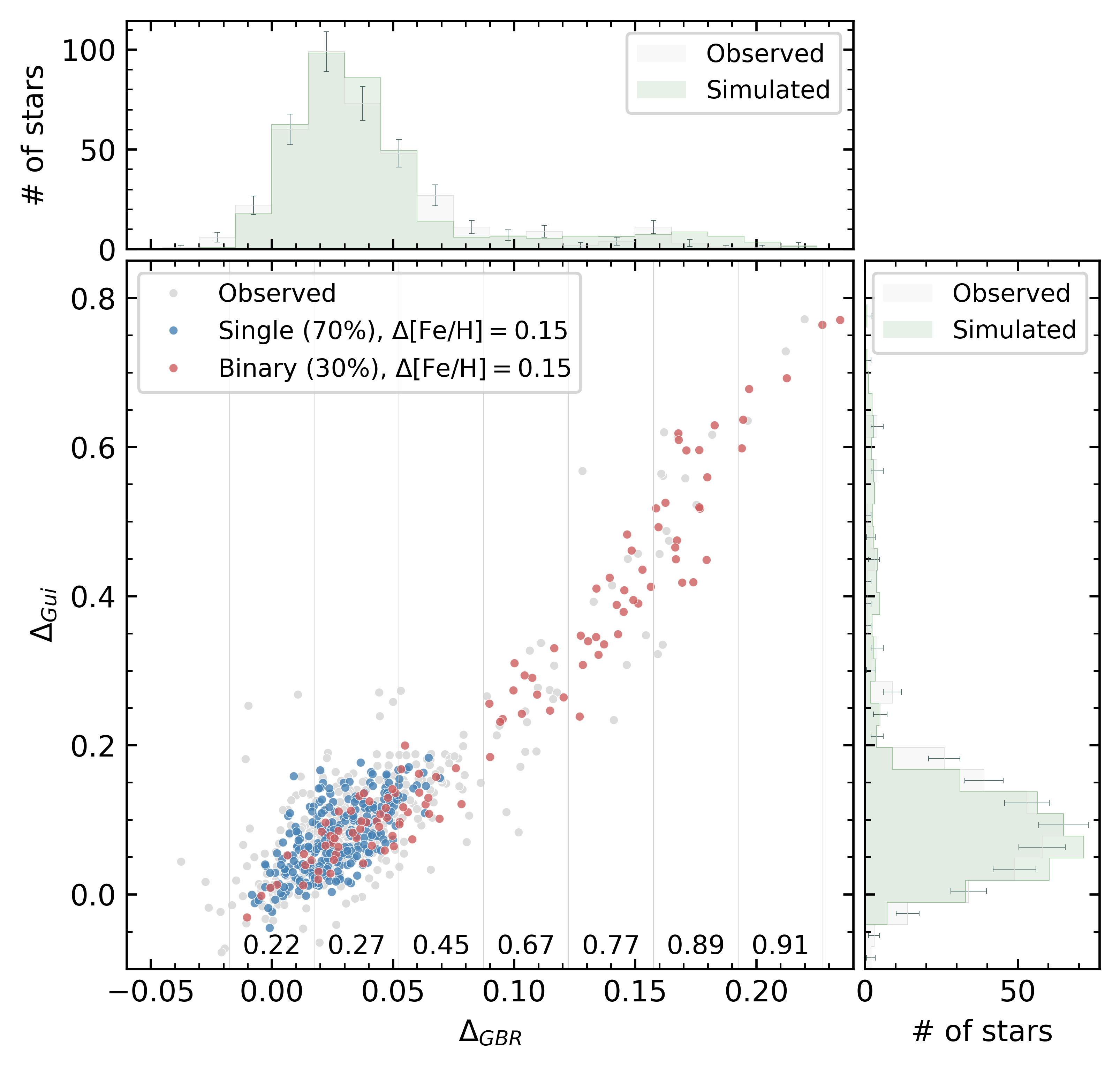}
        \caption{}
        \label{fig:z15}
     \end{subfigure}
     \hfill
     \begin{subfigure}{0.495\textwidth}
        \centering
        \includegraphics[width=\textwidth]{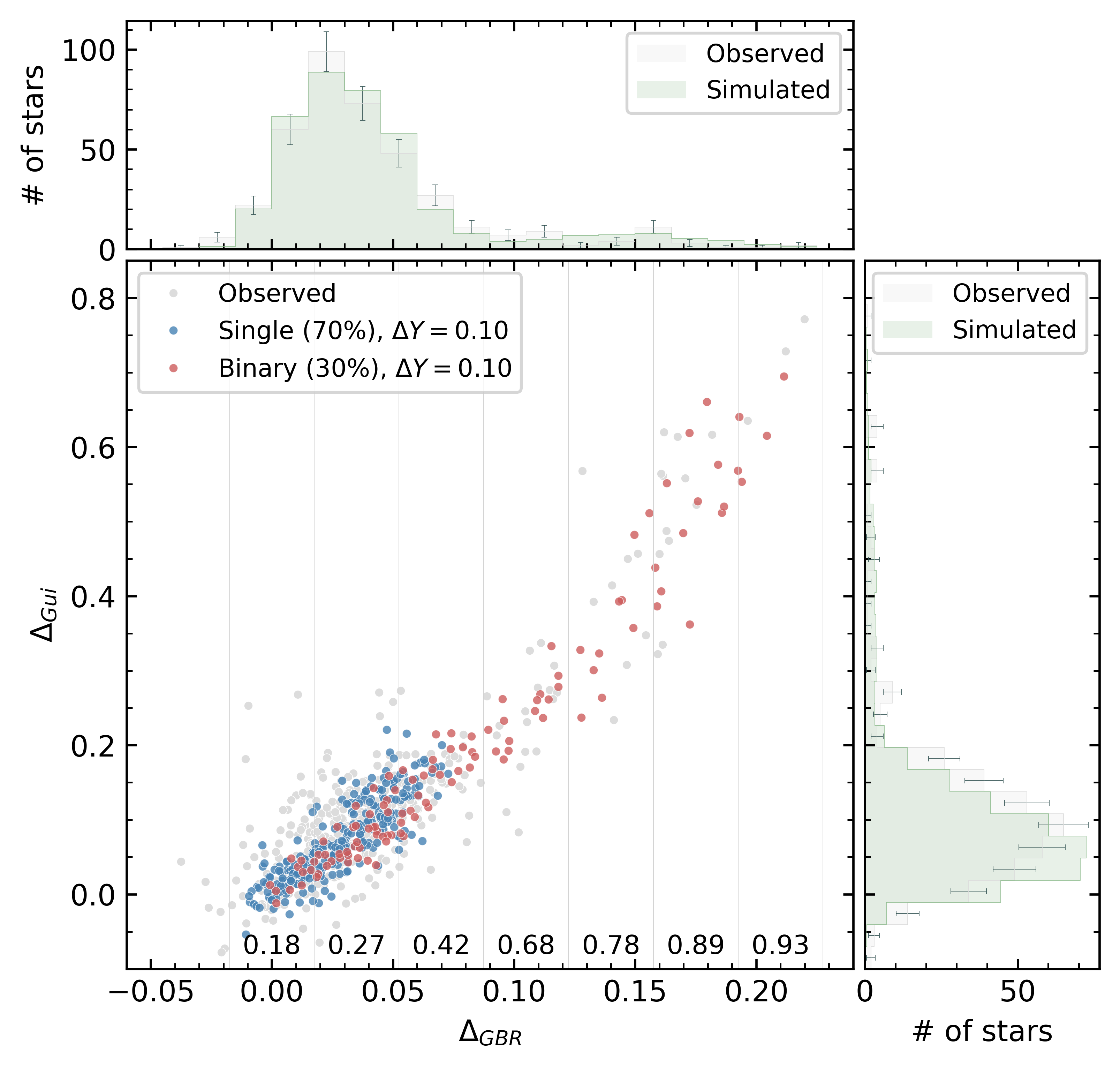}
        \caption{}
        \label{fig:he10}
     \end{subfigure}
        \caption{$\Delta_{GBR}$-$\Delta_{Gui}$ diagrams for the synthetic populations including binaries plus 
        a metallicity spread $\Delta \rm [Fe/H]=0.15$ (left) and a helium abundance spread $\Delta Y=0.10$ (right).
        The simulation in the left panel includes a reddening spread $\Delta E(B-V)=0.06$, while in the right panel we added a total amount amount of differential reddening $\Delta E(B-V)=0.03$.
        The values reported above the $x$-axis are the
        mean values of $q$ of the simulated binary stars falling in the regions delimited by the vertical thin lines.
        Observations are displayed as light grey filled circles. Synthetic single and unresolved binary stars are shown with different colours, to highlight their separate contributions to the diagram.
        On the right and at the top of each plot we show the histograms of the observed (with Poisson error bars on the star counts) and simulated number counts 
        along the two axes (see text for more details).}
        \label{fig:delta_delta}
\end{figure*}

We determined the $\Delta_{GBR}$-$\Delta_{Gui}$ diagram of this sample of unresolved binary stars 
with low $q$ values in the 
lower MS magnitude range defined before (which, as we have mentioned in the previous 
section, corresponds to a mass range of about 0.3\,$M_{\odot}$, and contains about 5500 objects in our simulation) by 
following the same procedure as for the cluster data (after applying distance modulus and reddening determined in the previous section), and fitted 
with a straight line reported in Fig.~\ref{fig:obs_red} the direction 
along which the synthetic stars move in this diagram due to the presence of unresolved companions.  Also in this case, the slope is shallower than observed. 

Given the inability of unresolved binaries and differential reddening 
to explain the trend displayed by the lower MS cluster stars in the $\Delta_{GBR}$-$\Delta_{Gui}$ diagram, we investigated also the effect 
of varying the initial chemical composition, namely the metallicity 
(parametrised in terms of [Fe/H]) and initial helium mass fraction $Y$.

The synthetic samples have been calculated as described before, but this time 
we assign to each mass a random value of [Fe/H] or $Y$ according 
to a uniform probability distribution with a range $\Delta \rm [Fe/H]=0.15$ (increasing 
from $\rm [Fe/H]=0.06$) or $\Delta Y=0.10$ (increasing from $Y=0.269$, the value of the BaSTI-IAC 
isochrones for $\rm [Fe/H]=0.06$, whose corresponding value of the 
metal mass fraction is kept constant in these simulations with varying $Y$) in case of variations of helium\footnote{In this analysis we consider $Y$ increasing above the reference $Y=0.269$. The reason is that a decrease of just $\Delta Y = 0.02$ would lead to the cosmological helium abundance, which seems unrealistic for a roughly Solar metallicity cluster.}.
An isochrone with the chosen chemical composition was first determined by interpolating 
quadratically among 600~Myr BaSTI-IAC isochrones of different [Fe/H] (and $Y$\footnote{To this purpose we have calculated additional isochrones with varying $Y$, metallicity $Z=0.0172$ and Solar scaled metal distribution, using the same 
code and physics inputs of the BaSTI-IAC models}), 
then the magnitudes (with added photometric errors) were determined as described before.

The direction of the synthetic sequences with [Fe/H] or $Y$ spreads are also 
reported in Fig.~\ref{fig:obs_red}. It is evident from the figure that the effect of a metallicity spread 
is predicted to move the stars along a steeper sequence compared to the observations 
(increasing [Fe/H] moves the objects towards larger values of both $\Delta_{GBR}$ and $\Delta_{Gui}$), whilst a 
spread of $Y$ moves the stars along almost the same direction of the observations (in this case objects with the highest $Y$ would display the lower values of $\Delta_{GBR}$ and $\Delta_{Gui}$ because an increase of $Y$ moves the MS towards the blue in both the $\textit{Gaia}$ and $Sloan$ CMDs).

The conclusion we can draw from the results in Fig.~\ref{fig:obs_red} is that 
the observed MS broadening can be explained potentially in two different ways.
The first possibility is a spread of metallicity among the cluster's stars, coupled with the presence of unresolved binaries (we know from the CMD that there are unresolved binaries with $q > 0.6$, hence 
there will be likely objects also with smaller $q$) and a range of $E(B-V)$ values --meaning differential reddening. 
These latter two effects tend to 
compensate for the too steep trend compared to the data predicted by just a metallicity spread.

The second possibility is a spread in $Y$, together with unresolved binaries and possibly a small 
amount of differential reddening, smaller than the case of a metallicity spread, otherwise the predicted trend 
in this diagram would become too shallow.

\begin{figure*}
    \centering
    \begin{subfigure}{0.495\textwidth}
        \centering
        \includegraphics[width=\textwidth]{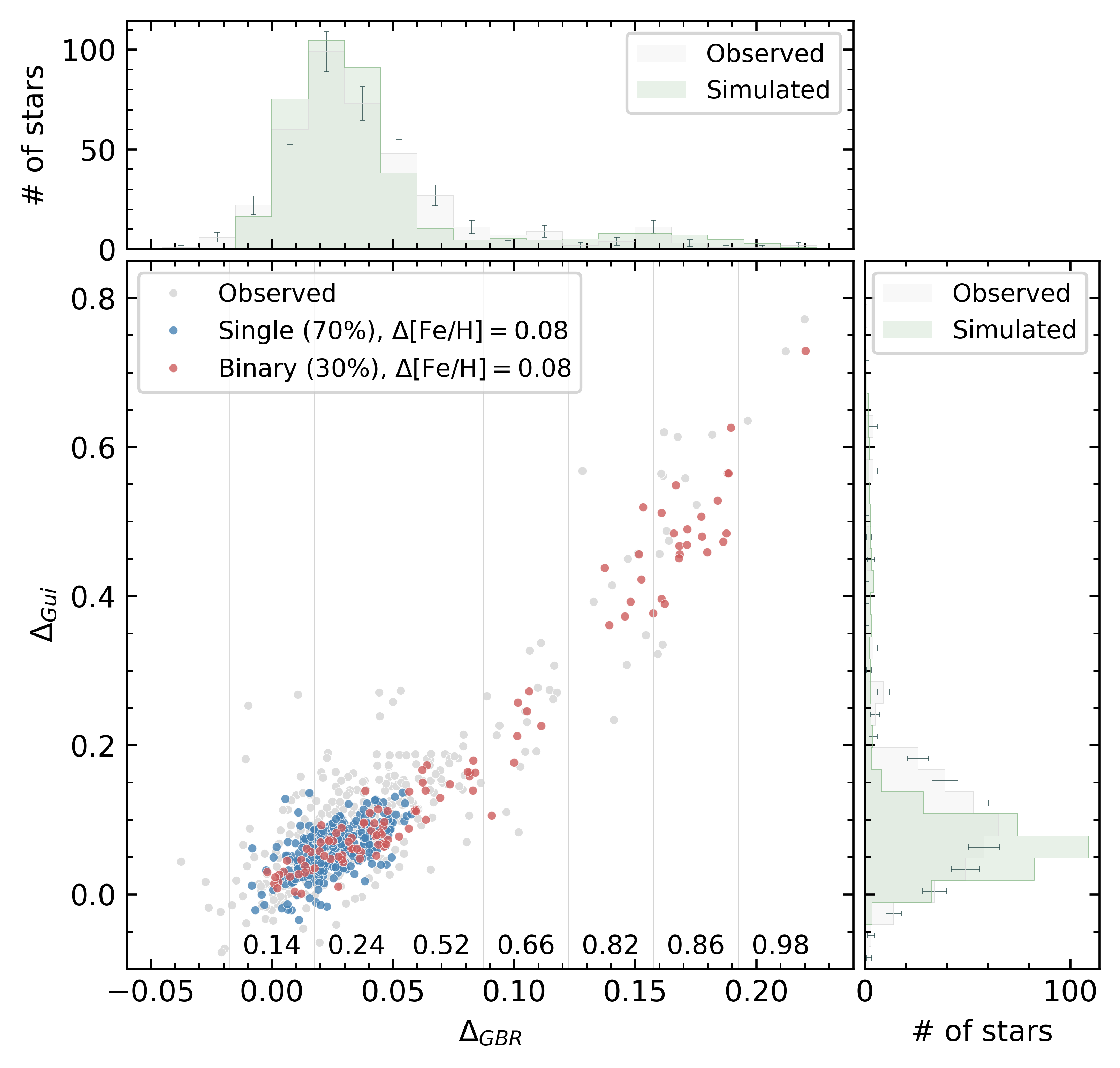}
        \caption{}
        \label{fig:z08}
    \end{subfigure}
    \hfill
    \centering
    \begin{subfigure}{0.495\textwidth}
        \centering
        \includegraphics[width=\textwidth]{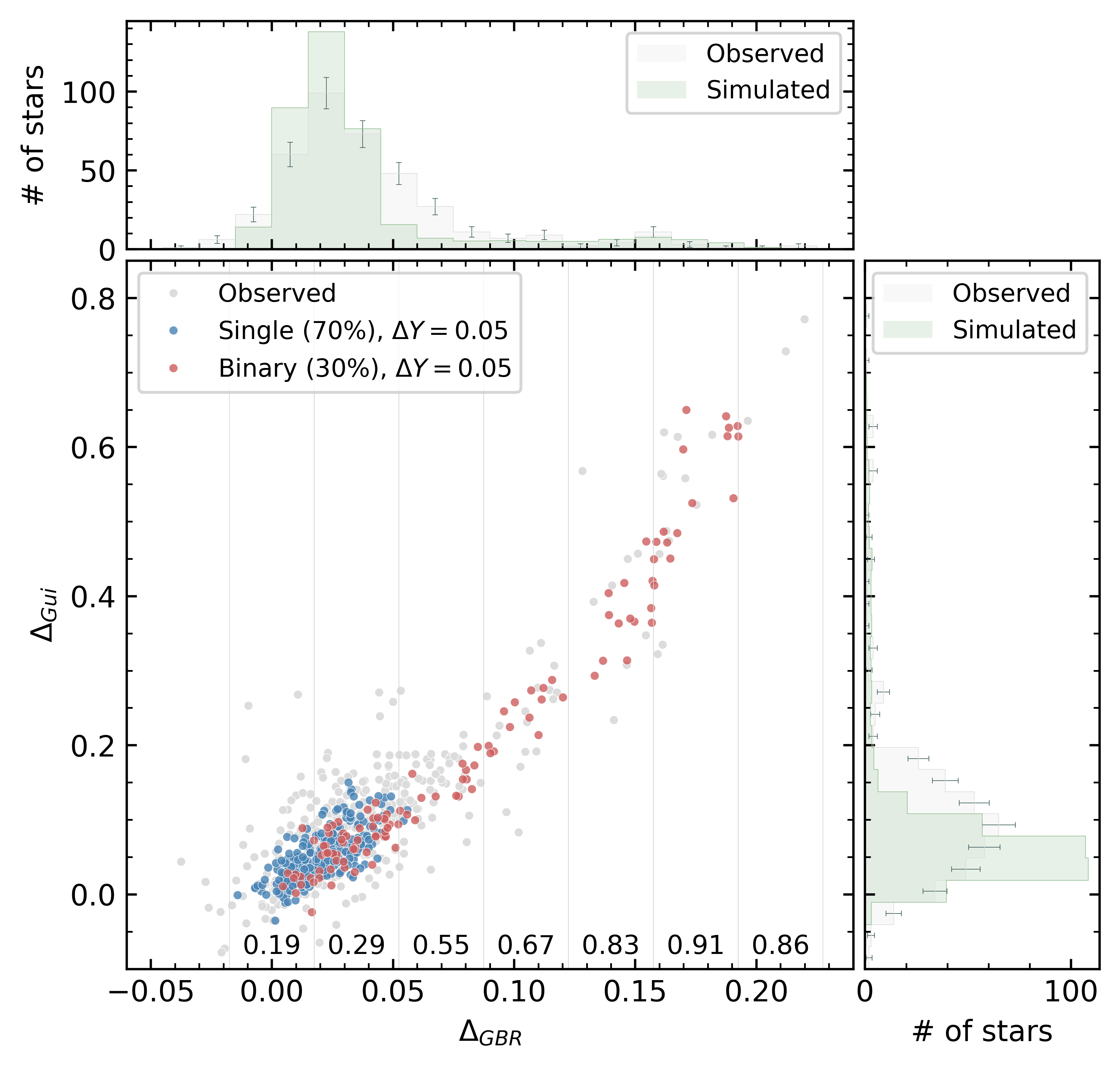}
        \caption{}
        \label{fig:he05}
    \end{subfigure}
    \caption{As Fig.\,\ref{fig:delta_delta}, but with $\Delta \rm [Fe/H]=0.08$ (left) 
    and $\Delta Y =0.05$ (right).}
    \label{fig:delta_delta_1}
\end{figure*}

To set constraints on the size of the metallicity and helium spreads, and the amount of differential reddening for these two scenarios, we have performed additional simulations as those just described, by varying $\Delta$[Fe/H], $\Delta Y$ and $\Delta E(B-V)$, to reproduce the observed distribution of points in the $\Delta_{GBR}$-$\Delta_{Gui}$ diagram.
In all these simulations we assumed a binary fraction equal to 0.30, consistent with the value determined by \citet{cordoni18} considering the fraction of unresolved binaries with $q>0.7$ determined from the CMD,  
and a flat probability distribution of the values of the mass ratios\footnote{We also considered the case of a power law distribution of $q$ (probability distribution proportional to $q^{-0.6}$) as found by \citet{malofeeva} for the Pleiades. In this case, the fraction of unresolved binaries with $q>0.7$ determined by \citet{cordoni18} provides a total binary fraction equal to $\sim0.7$. The constraints on the values of $\Delta$[Fe/H], $\Delta Y$ and $\Delta E(B-V)$ however do not change compared to our reference simulations. The reason is that 
most of this increase of the number of binaries happens for low values of $q$, and in this situation the magnitudes and colours of the binaries are almost coincident with those of the primary component, hence of single stars.}.

Given that we do not know the real probability distribution of the values of reddening and [Fe/H] or $Y$ in the observed sample of stars, we stick to a uniform distribution. This means that in principle we cannot expect to find a perfect match to the observations in the $\Delta_{GBR}$-$\Delta_{Gui}$ diagram. However, we can still set important constraints on the size of these spreads by trying to simultaneously reproduce as best as possible the number distributions along the two axes of this diagram. 

Figure~\ref{fig:delta_delta} shows two simulations compared to observations 
in the $\Delta_{GBR}$-$\Delta_{Gui}$ diagram, including also objects compatible 
with being binaries with high values of $q$.
The synthetic samples contain about 5500 objects and include observational errors, unresolved binaries, differential reddening --a spread 
of $E(B-V)$-- and either a spread of [Fe/H] or a spread of the initial helium mass fraction. In the figure we show only one random subset of the full sample, which contains the same number of objects as the observations. 
The histograms along the horizontal and vertical axis compare the number distributions of the synthetic and real stars as a function of $\Delta_{GBR}$ and $\Delta_{Gui}$, 
respectively. In this case, we have considered the full sample of synthetic stars, and rescaled the histograms to have the same total number of objects as observed. This way we minimize the Poisson error on the number counts for the synthetic sample.
The bin size of the 
histograms is about two times the average $1 \sigma$ error bars on 
$\Delta_{GBR}$ and $\Delta_{Gui}$ over the $G$ magnitude range of the observed sample.

These simulations have been performed considering $\Delta \rm [Fe/H]=0.15$, $\Delta Y=0.10$, and $\Delta E(B-V)=0.06$ and 0.03 for the case of metallicity and helium spread, respectively, and provide a general satisfactory agreement with the distribution of the observed stars in this diagram. 
The synthetic samples cover nicely the region of the diagram populated by the cluster stars, even when considering the binaries with high $q$ values. Also the number distributions along the two axes 
are reasonably consistent with the data, within the error bars on the star counts. 

Of course, combinations like $\Delta \rm [Fe/H]=0.16$ together with $\Delta E(B-V)=0.07$, or 
$\Delta Y=0.10$ with $\Delta E(B-V)=0.01$, cannot be excluded from this kind of comparisons, 
but values of $\Delta$[Fe/H], $\Delta Y$, and $\Delta E(B-V)$ very different from those chosen 
for the simulations in Fig.~\ref{fig:delta_delta} can indeed be discarded, as shown below.

\begin{figure*}
    \centering
    \begin{subfigure}{0.495\textwidth}
        \centering
        \includegraphics[width=\textwidth]{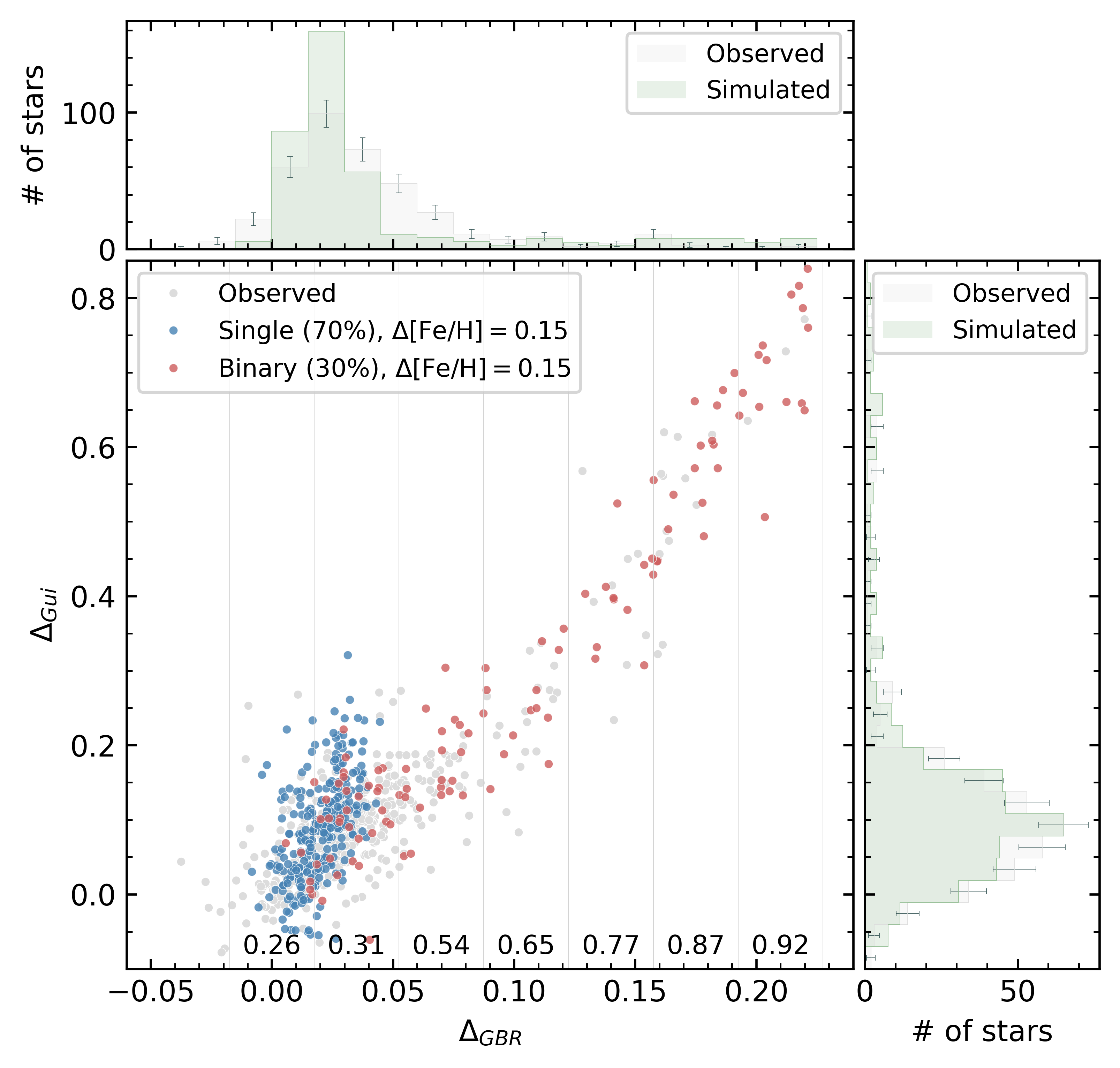}
        \caption{}
        \label{fig:z15red00}
    \end{subfigure}
    \centering
    \begin{subfigure}{0.495\textwidth}
        \centering
        \includegraphics[width=\textwidth]{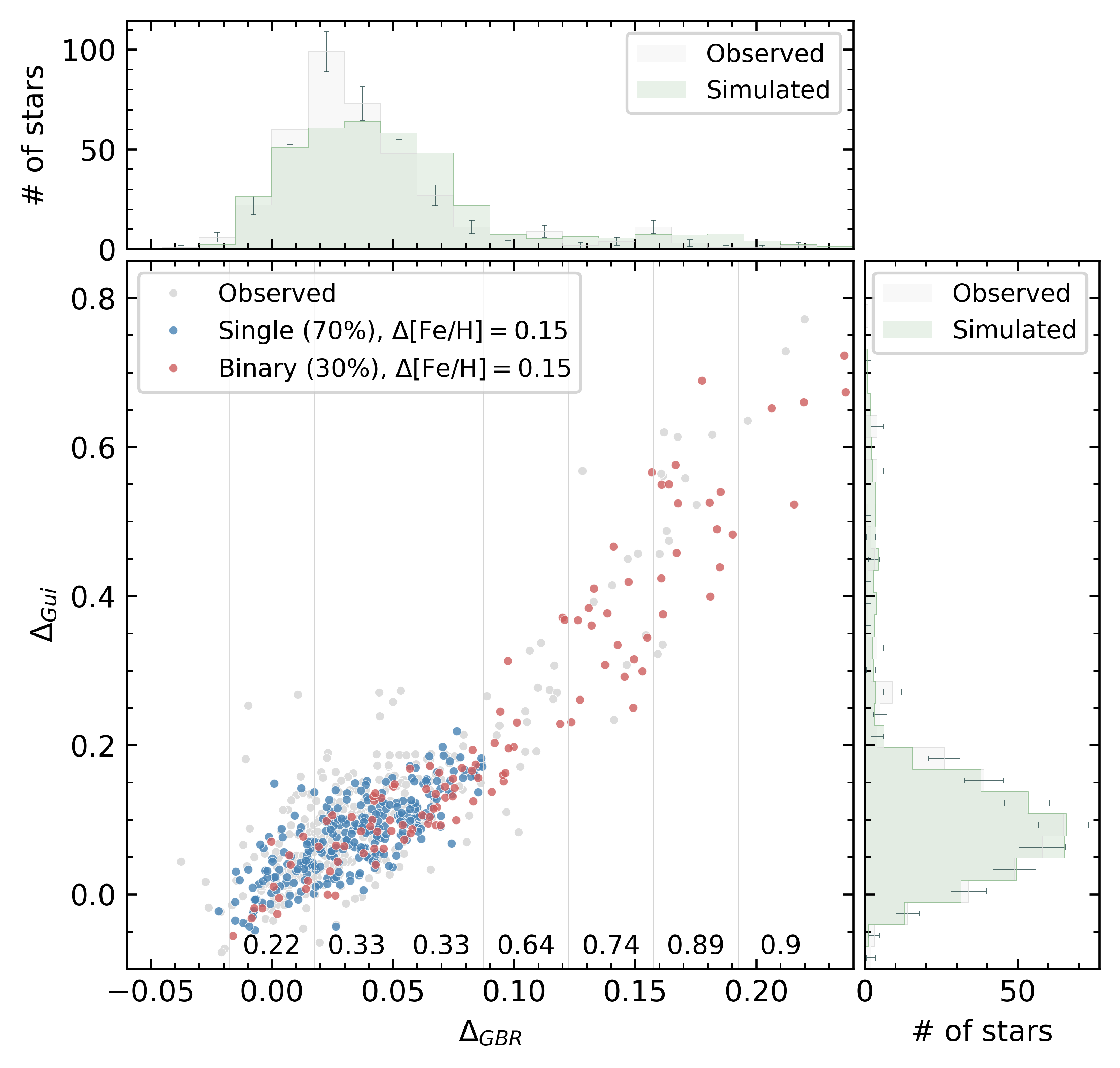}
        \caption{}
        \label{fig:z15red01}
    \end{subfigure}
    \hfill
    \centering
    \begin{subfigure}{0.495\textwidth}
        \centering
        \includegraphics[width=\textwidth]{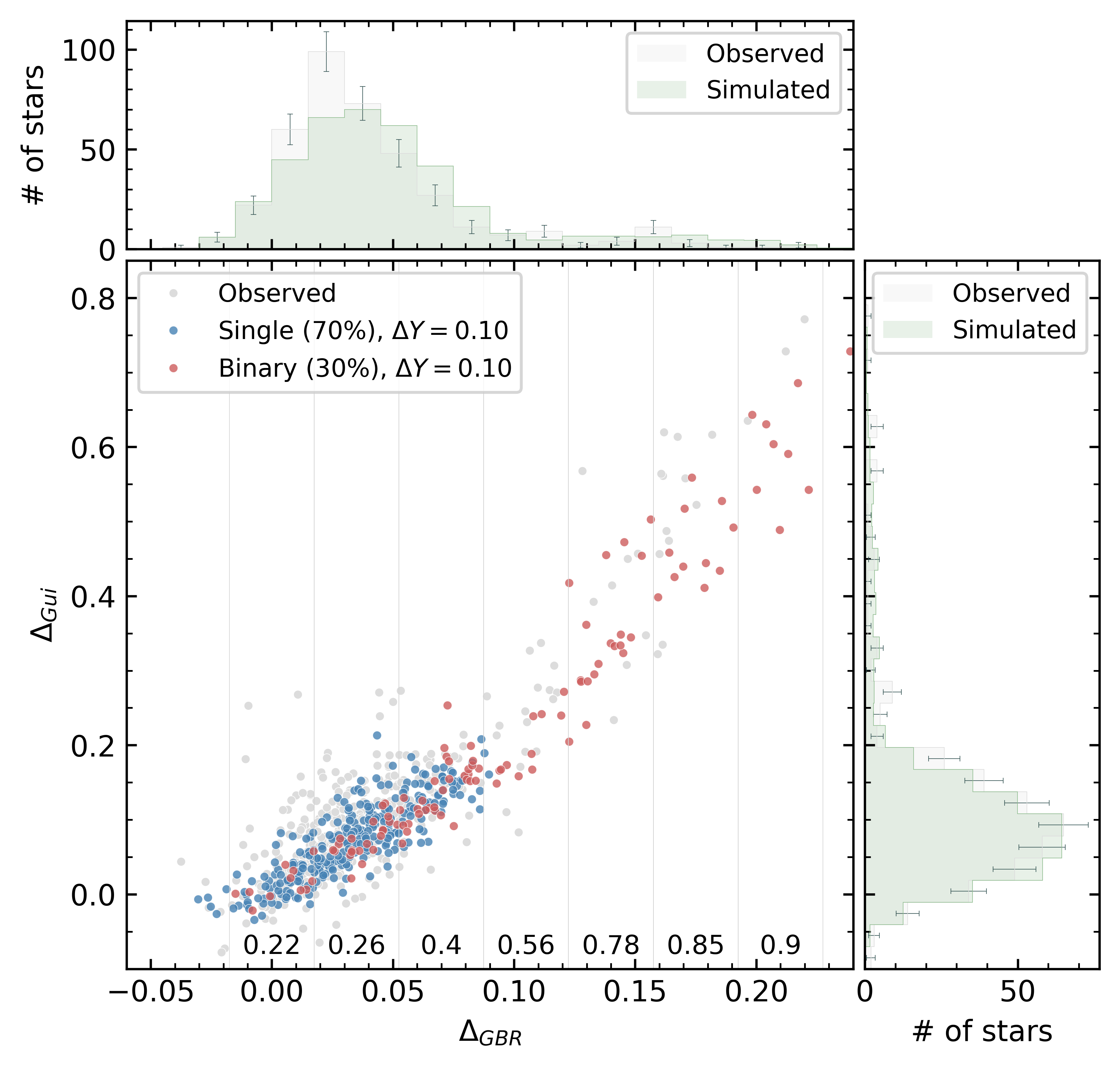}
        \caption{}
        \label{fig:he10red08}
    \end{subfigure}
    \caption{As Fig.~\ref{fig:delta_delta}, but with $\Delta \rm [Fe/H]=0.15$ and $\Delta E(B-V)=0$ (a), $\Delta \rm [Fe/H]=0.15$ and $\Delta E(B-V)=0.10$ (b),  $\Delta Y=0.10$ and $\Delta E(B-V)=0.08$ (c).}
    \label{fig:delta_delta_red}
\end{figure*}  

Figure~\ref{fig:delta_delta_1} compares the data with 
two simulations like those in Fig.~\ref{fig:delta_delta}, keeping the 
$\Delta E(B-V)$ values unchanged, but reducing the spread   
$\Delta \rm [Fe/H]$ to 0.08~dex and $\Delta Y$ to 0.05. The agreement with the 
observations is in this case much worse. In both cases the synthetic samples are clearly short of stars in the region with $\Delta_{GBR}$ between $\sim 0.06$ and $\sim 0.08$~mag, and $\Delta_{Gui}$ between $\sim 0.10$ and $\sim 0.2$~mag.
The synthetic single stars do not reach that area and 
changing the probability distribution of $q$ to a power law 
\citep[][with the consequent increase of the overall binary fraction]{malofeeva} simply replaces many of the current single stars with unresolved binaries, but it does not help populating that region of the diagram. 
The opposite would happen if we increase $\Delta \rm [Fe/H]$ and $\Delta Y$ well above the values in Fig.~\ref{fig:delta_delta}, with the bulk of the synthetic sequences (essentially the single star sequence) 
too extended towards higher values of $\Delta_{GBR}$ and $\Delta_{Gui}$ compared to the observations. 

Figure~\ref{fig:delta_delta_red} displays the effect of varying $\Delta E(B-V)$ in the simulations, 
by keeping fixed $\Delta \rm [Fe/H]$ and $\Delta Y$ to the values of the simulations in Fig.~\ref{fig:delta_delta} ($\Delta \rm [Fe/H]=0.15$, $\Delta Y=0.10$).
We can see here that a smaller amount of differential reddening for the 
case with metallicity spread makes the single star sequence too steep compared to the observations, 
consistently with the results in Fig.~\ref{fig:obs_red}, 
whilst a larger $\Delta E(B-V)$ makes the overall slope of the synthetic sequence shallower than the observations. This can be clearly appreciated 
in the figure when looking at the region populated by binaries with high $q$.
In case of the helium abundance spread, increasing $\Delta E(B-V)$ makes again 
the whole sequence shallower, again consistent with the results of Fig.~\ref{fig:obs_red}.

Finally, we have repeated this whole analysis by considering 
for each star the two completely independent 
CMDs $G$-$(G_{\mathrm{BP}}-G_{\mathrm{RP}})$ and $g$-$(u-i)$. From these CMDs we 
have calculated 
$\Delta_{GBR}$ as described before, and $\Delta_{gui}$, and 
compared data with simulations in $\Delta_{GBR}$-$\Delta_{gui}$ diagrams. 
The quantity $\Delta_{gui}$ is analogous to 
$\Delta_{Gui}$, but this time the difference in $(u-i)$ colour is taken 
with respect to the corresponding value of the blue fiducial in the 
$g$-$(u-i)$ CMDs at the star $g$ magnitude.
The conclusions are exactly the same as when using $\Delta_{GBR}$-$\Delta_{Gui}$ diagrams, as shown by the figures included in the Appendix \ref{sec:appx}.

%%%%%%%%%%%%%%%%%%%%%%%%%%%%%%%%%%%%%%%%%%%%%%%%
%%%%%%%%%%%%%%%%%%%%%%%%%%%%%%%%%%%%%%%%%%%%%%%%
%%%%%%%%%%%%%%%%%%%%%%%%%%%%%%%%%%%%%%%%%%%%%%%%
%%%%%%%%%%%%%%%%%%%%%%%%%%%%%%%%%%%%%%%%%%%%%%%%
\section{Discussion and conclusions}

The very accurate {\it Gaia} EDR3 photometry for the members of M37 presented by \citet{gb2022} shows 
a MS broadened in colour beyond what expected from photometric errors only, well below the region of the extended TO, where neither age differences nor rotation are expected to play a role. Even when we neglected redder objects compatible with being unresolved binaries with mass ratios above $q \sim 0.6$-0.7, the cluster MS is still broader than expected from the 
small photometric errors.

To investigate the causes of this broadening we made use of an auxiliary photometry in the {\it Sloan} system,
and  built a differential colour-colour diagram of the lower MS, using a combination of {\it Gaia} and {\it Sloan} filters.
By employing synthetic stellar populations to reproduce the observed trend of the cluster stars in this diagram, we have 
concluded that the observed colour spread in the {\it Gaia} CMD can be reproduced by a combination of either a metallicity spread $\Delta \rm[Fe/H] \sim 0.15$ plus a differential reddening across the face of the cluster spanning a total 
range $\Delta E(B-V) \sim 0.06$, or 
an initial helium abundance spread $\Delta Y\sim0.10$ plus a smaller range of reddening $\Delta E(B-V) \sim 0.03$.

\begin{figure}
    \centering
    \includegraphics[width=\columnwidth]{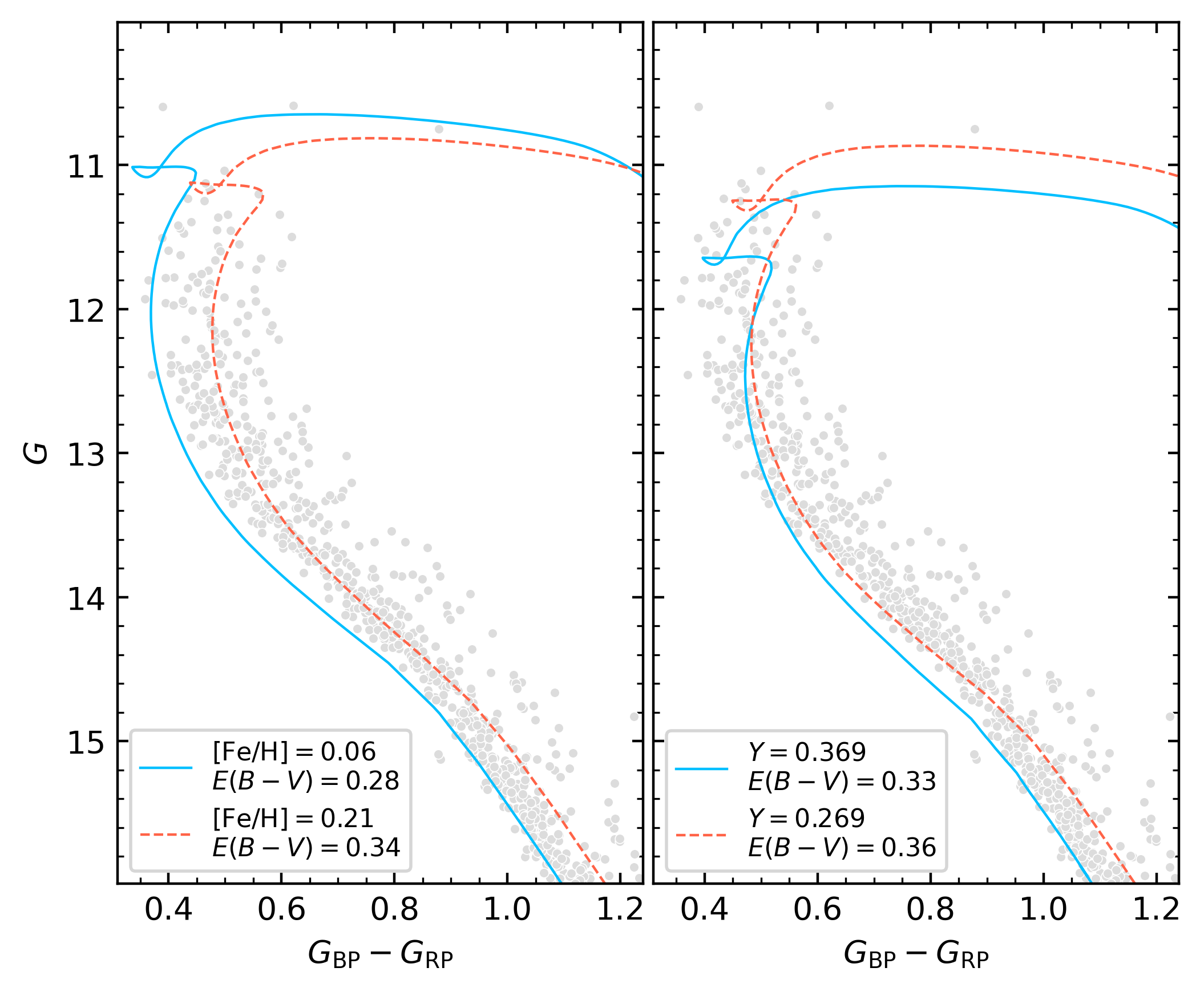}
    \caption{Effect of the  metallicity plus reddening spread (left) and $Y$ plus reddening spreads (right) scenarios on the TO region of the cluster {\it Gaia} CMD. We show 380~Myr isochrones 
    with the combinations of [Fe/H], $Y$ and $E(B-V)$ displayed in the two panels (see text for details).}
    \label{fig:cmd_iso_diff}
\end{figure}

Figure~\ref{fig:cmd_iso_diff} shows the impact of these two scenarios on the TO region of 
the cluster's population. 
We display in the two panels a 380~Myr isochrone with either different 
combinations of metallicity and reddening (left panel) or different 
combinations of $Y$ and reddening, keeping the metallicity fixed (right panel).
In the reasonable assumption that there is no correlation between 
reddening and chemical composition, the combinations we display match the 
blue and red limits of the single star sequence in the $G$ magnitude range 
we have studied.
Notice that in case of a helium abundance spread the smallest value of  
$E(B-V)$ is higher than that for the case of a metallicity spread. The reason is that an 
increase of the helium abundance shifts the MS to the blue, 
hence in this scenario a match of the blue edge of the observed CMD with the bluest stellar component (the population  with the highest helium and lowest reddening) requires a  value of $E(B-V)$ larger than that determined from the fit in Fig.~\ref{fig:cmd_iso}, and used in the left panel for the case of a metallicity spread.

The metallicity (and reddening) spread has a major impact around the TO, 
with single age non rotating isochrones able to cover in this case a large 
portion of the extended TO of the cluster CMD. The impact at the TO is much less pronounced for the case of a helium (and reddening) spread.

Another consequence of these abundance spreads is that we need 
to be very cautious when applying differential reddening corrections to the cluster CMD. 
In fact, none of the CMDs displayed in our study has been corrected for this effect.

Nowadays it is a standard procedure to correct for differential reddening the CMDs of star clusters 
\citep[e.g.,][]{2007AJ....133.1658S,diffred,2017ApJ...842....7B}; the basic physical principles
underlining the method are summarised in the following.

First, stars along a portion of the cluster MS are selected as \lq{reference stars\rq}, 
and a reference fiducial line for these stars is calculated, by determining a median colour at varying magnitude. The magnitude range of this reference sequence of stars is chosen such that the 
direction of the reddening vector can be more easily discriminated from the effect of photometric errors.
To a generic target star in any evolutionary phase at a
given spatial position within the cluster is then assigned a number of neighbouring (spatially) reference stars, and the median distance along the reddening vector between the position of 
these neighbouring reference stars and the reference fiducial line 
is calculated (this distance can be then transformed into a value of $\Delta E(B-V)$). 
The position in the CMD of the target star is then shifted along the reddening vector by 
the value of this median distance 
and the procedure is repeated for all cluster stars, 
including the individual reference objects.

An underlying assumption of this method is that the cluster stars are a homogeneous stellar population, but it 
could still work in case of chemical inhomogeneities, provided that two conditions are satisfied. 
The first one is that the chemical properties of the stellar population are not spatially dependent (the size and probability distribution of the abundance spreads are not dependent on 
the location within the cluster). The second one is that for any object there is a sufficiently large number of neighbouring reference stars, such that they properly sample the full range of 
chemical inhomogeneities. This way they would define a sequence in the CMD that is equivalent 
(in terms of stellar properties) to the fiducial reference line.

If we were to be able to determine with confidence 
the amount of differential reddening in our CMD, we might possibly discriminate between the 
metallicity and helium spread scenarios, because they require different values of 
$\Delta E(B-V)$. Unfortunately, we do not have enough reference stars to do that. The best 
portion of the MS to use for the reference fiducial line is in the range between $G \sim 15$
and $\sim 17$, and by following the technique described before, we find that 
for objects in large portions of the cluster we have only 10 or less neighbouring
reference stars (within a circle of radius $\sim 3$\,arcmin).
With this method we get a total range $\Delta E(B-V)$ across the cluster equal to 
$\sim 0.06$-0.07~mag supporting the scenario with a metallicity spread, 
but with such small numbers of local reference stars we 
cannot be sure they are sufficient to properly sample the  
distribution of chemical abundances, hence we might have biases in the estimate of the local 
differential reddening. 

To discriminate more reliably between metallicity and helium spread, high-resolution differential abundance determinations of a sizeable sample of cluster stars are then necessary, 
because they can confirm or exclude the presence of a metal abundance spread.
The existing more direct measurements --based very small samples of targets-- do not allow to draw solid conclusions. 
\citet{marshall} published  moderate resolution spectroscopy of a sample of eight red clump stars in the cluster, and found 
that the derived metallicities of the target-cluster stars displayed 
a scatter of 0.14~dex, about twice what expected from measurement errors.
On the other hand, the high-resolution spectroscopy of three red clump stars by \citet{pancino} does not reveal any clear abundance spread. 

Irrespective of the uncertainty between metallicity and helium spread, our results raise the possibility that also open clusters --like globular clusters and in general massive 
star clusters \citep[see, .e.g.][and references therein]{gratton, bl, martocchia, marino, legnardi, lscb}-- do not host stars all with the same initial chemical composition. So far, there has been some debate about the presence of a metallicity spread in the open cluster Tombaugh~2, with the 
high-resolution spectroscopy by  
\citet{frinchaboy} who found the presence of a metallicity spread in the cluster, that was however not confirmed by the subsequent spectroscopic analysis by \citet{villanova}. Our work adds a new candidate open cluster hosting chemical abundance spreads.

\section*{Acknowledgements}
Based on observations collected at the Schmidt telescope (Asiago, Italy) of INAF.
This work has made use of data from the European Space Agency (ESA) mission
{\it Gaia} (\url{https://www.cosmos.esa.int/gaia}), processed by the {\it Gaia}
Data Processing and Analysis Consortium (DPAC,
\url{https://www.cosmos.esa.int/web/gaia/dpac/consortium}). Funding for the DPAC
has been provided by national institutions, in particular the institutions
participating in the {\it Gaia} Multilateral Agreement.

This research or product makes use of public auxiliary data provided by ESA/Gaia/DPAC/CU5 and prepared by Carine Babusiaux.

MG and LRB acknowledge support by MIUR under PRIN program \#2017Z2HSMF. 
MS acknowledges support from The Science and
Technology Facilities Council Consolidated Grant ST/V00087X/1.
SC acknowledges financial support from Premiale INAF MITiC, from INFN (Iniziativa specifica TAsP), and from
PLATO ASI-INAF agreement n.2015-019-R.1-2018.

%%%%%%%%%%%%%%%%%%%%%%%%%%%%%%%%%%%%%%%%%%%%%%%%%%
\section*{Data Availability}

The isochrones employed in this study can be retrieved at \url{http://basti-iac.oa-abruzzo.inaf.it}, but for the helium enhanced isochrones, that are available upon request. 

The calibrated photometry and astrometry employed in this article is 
the one presented in \cite{griggio22},
which is also released
as supplementary on-line material, and available at this 
\url{https://web.oapd.inaf.it/bedin/files/PAPERs\_eMATERIALs/M37\_ugiSchmidt/}, 
along with an atlas.
The same catalogue also conveniently lists the \textit{Gaia}\,EDR3 photometry, astrometry and 
source ID, when available \citep{2021A&A...649A...1G}.



\bibliographystyle{mnras}
\bibliography{bibliography} % if your bibtex file is called example.bib

%\begin{thebibliography}{99}
%\end{thebibliography}

%%%%%%%%%%%%%%%%%%%%%%%%%%%%%%%%%%%%%%%%%%%%%%%%%%

%%%%%%%%%%%%%%%%% APPENDICES %%%%%%%%%%%%%%%%%%%%%

\appendix

\section{Complementary analysis}
\label{sec:appx}

\begin{figure*}
    \centering
    \begin{subfigure}{0.495\textwidth}
        \centering
        \includegraphics[width=\textwidth]{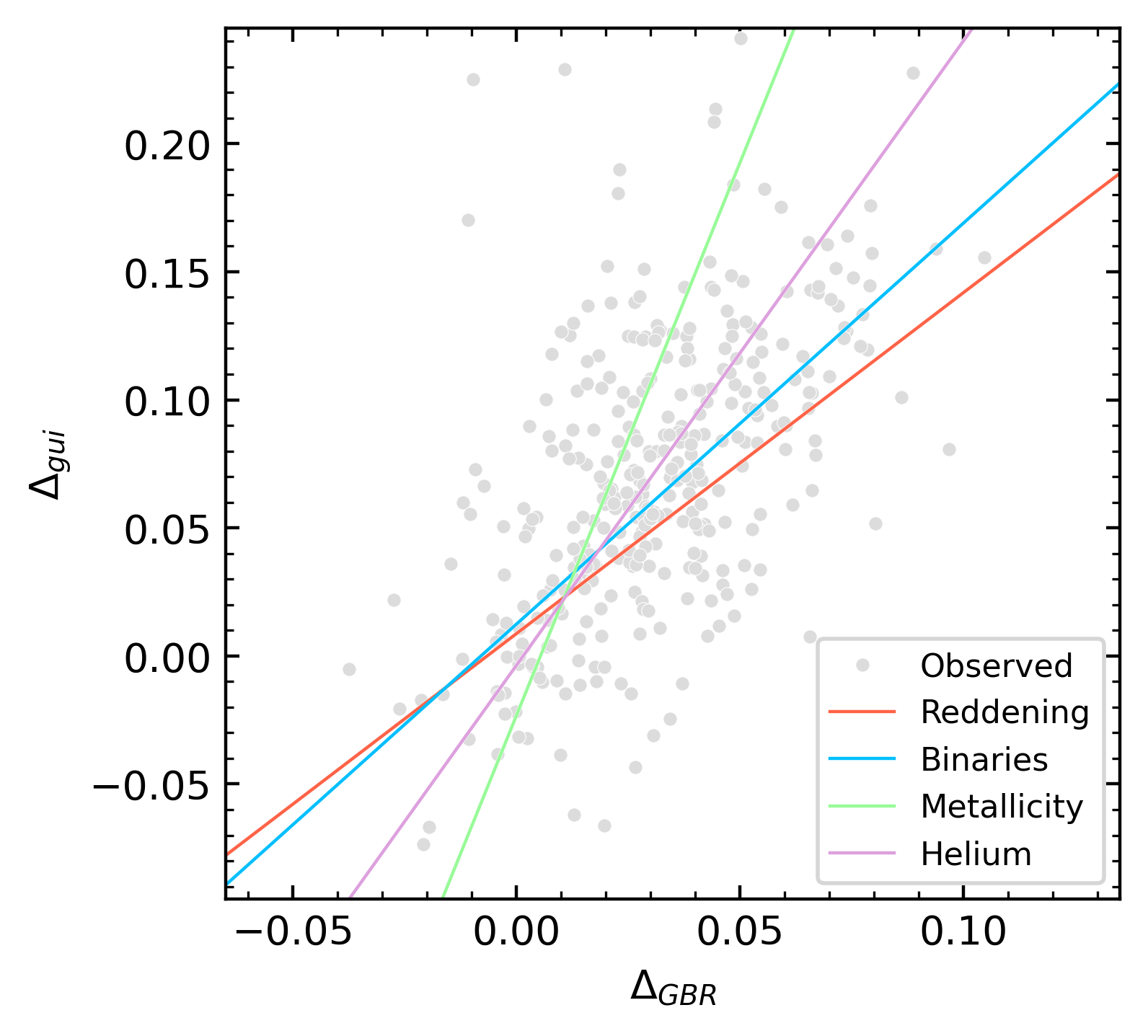}
        \caption{}
        \label{fig:obs_red_bis}
    \end{subfigure}

    \centering
    \begin{subfigure}{0.495\textwidth}
        \centering
        \includegraphics[width=\textwidth]{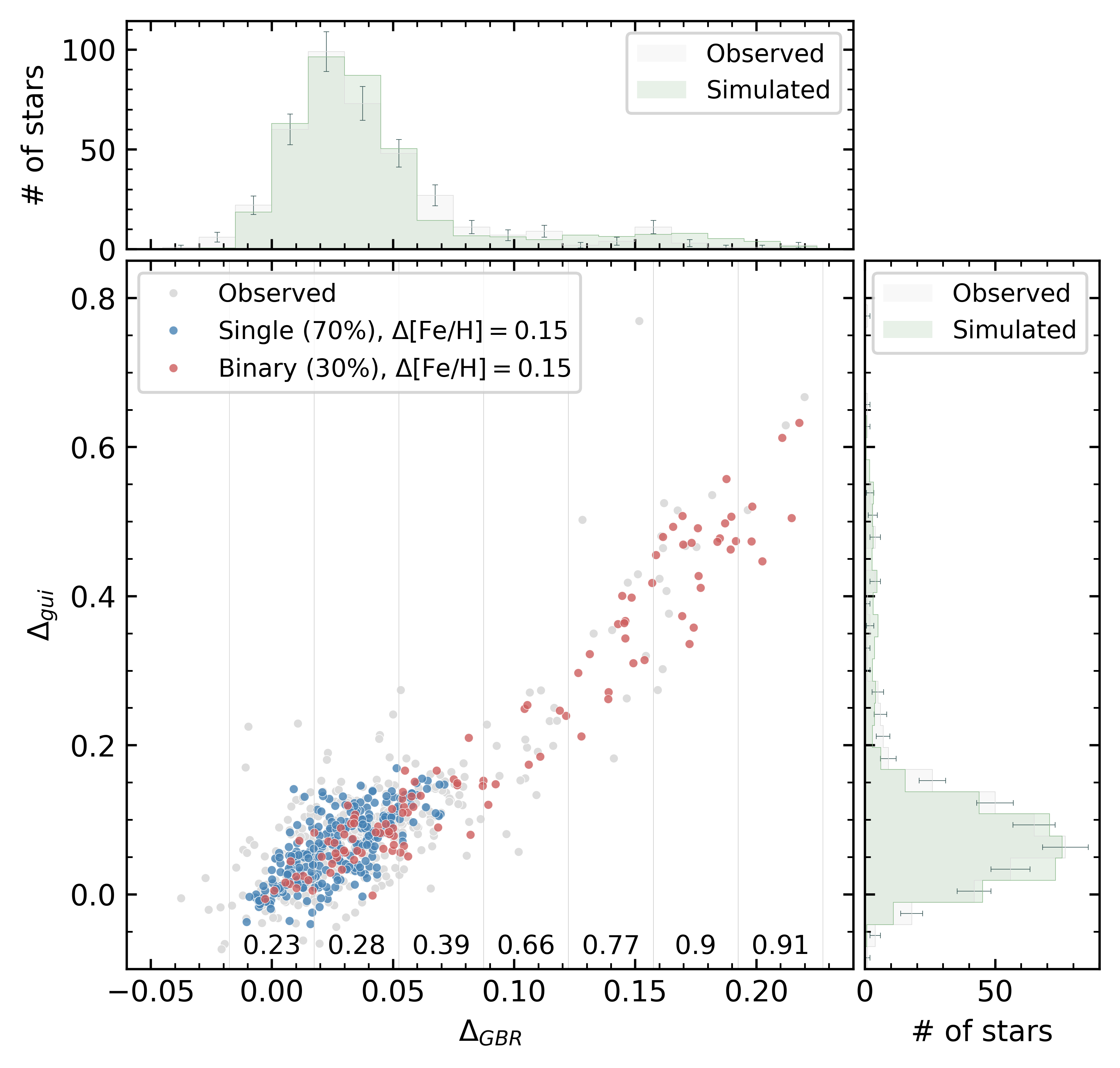}
        \caption{}
        \label{fig:z15_bis}
    \end{subfigure}
    \hfill
    \begin{subfigure}{0.495\textwidth}
        \centering
        \includegraphics[width=\textwidth]{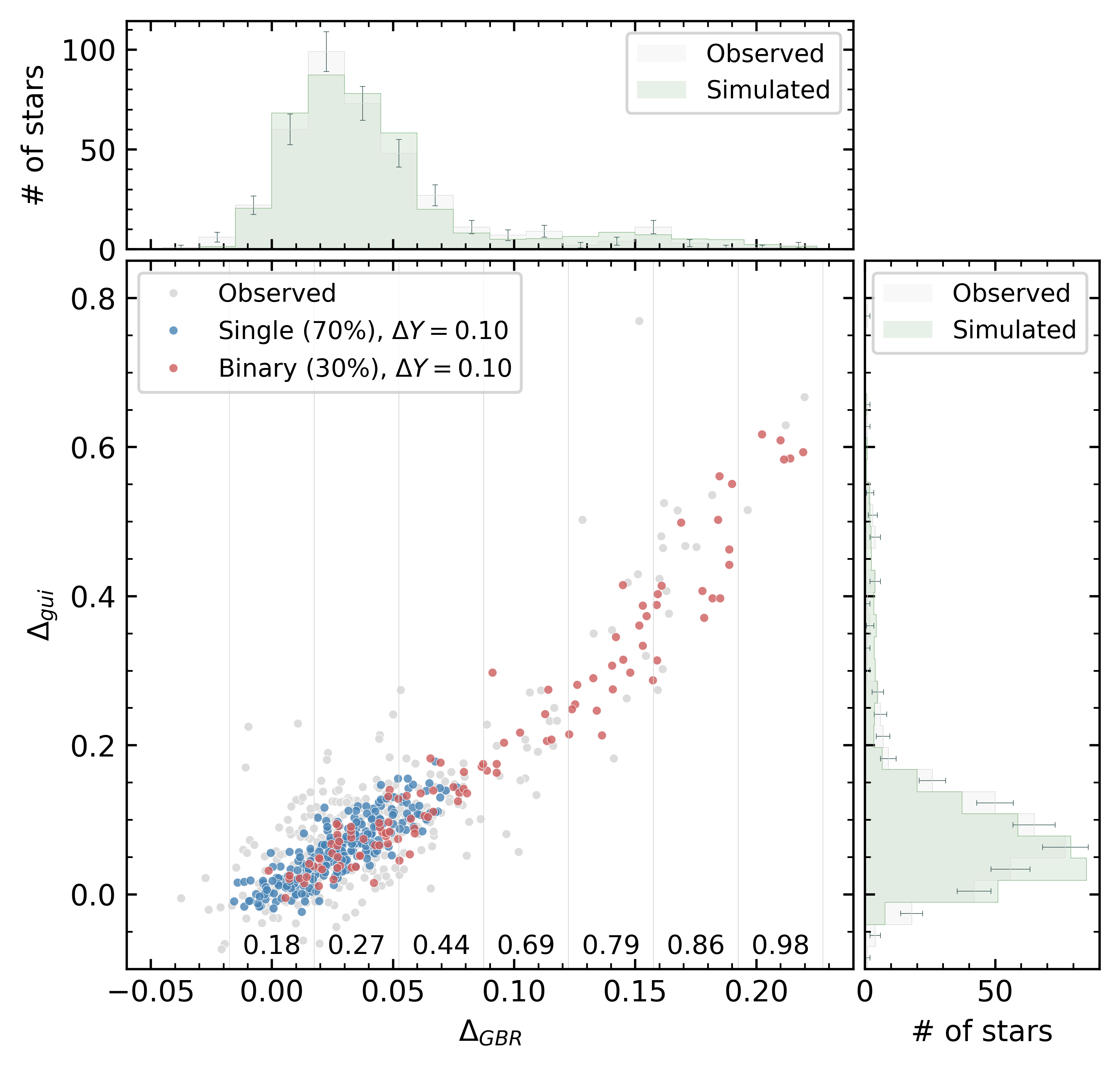}
        \caption{}
        \label{fig:he10_bis}
    \end{subfigure}
    \caption{ Panel (a): As Fig.~\ref{fig:obs_red} but in the 
         $\Delta_{GBR}$-$\Delta_{gui}$ diagram.
         Panels (b) and (c): As Fig.~\ref{fig:delta_delta} but in the 
         $\Delta_{GBR}$-$\Delta_{gui}$ diagram.}
    \label{fig:delta_delta_bis}
\end{figure*}

We show here some results of the same analysis described in Sect.~\ref{broad}, but 
using the $\Delta_{GBR}$-$\Delta_{gui}$ diagrams instead of $\Delta_{GBR}$-$\Delta_{Gui}$ ones. 
Figure~\ref{fig:obs_red_bis} is the equivalent of Figure~\ref{fig:obs_red}, 
and displays the lower MS stars, after excluding sources whose colours are consistent with unresolved binaries with mass ratio $q\gtrsim 0.6$. 
The four straight lines display the direction along which stars are displaced 
due to the effects of differential reddening, unresolved binaries with $q\lesssim 0.6$, spread of initial metal content, and helium abundance, respectively.

Figures~\ref{fig:z15_bis} and \ref{fig:he10_bis} are the equivalent of Fig.~\ref{fig:delta_delta} 
and display the lower MS stars including unresolved binaries with high $q$ values, 
together with synthetic stars calculated including a spread 
in metallicity $\Delta \rm [Fe/H]=0.15$ and reddening $\Delta E(B-V)=0.06$ (b), 
and a spread of initial helium $\Delta Y=0.10$ and reddening $\Delta E(B-V)=0.03$ (c).

% Don't change these lines
\bsp	% typesetting comment
\label{lastpage}
\end{document}